\documentclass[aps,pra,twocolumn,showpacs,superscriptaddress]{revtex4}
\usepackage{graphicx}

\def\be{\begin{equation}}       \def\ee{\end{equation}}
\def\bea{\begin{eqnarray}}      \def\eea{\end{eqnarray}}

\begin{document}

\title{Spin waves in the $(\pi,0)$ magnetically ordered iron chalcogenide Fe$_{1.05}$Te}

\author{O.J. Lipscombe}
\affiliation{The University of Tennessee, Knoxville,
Tennessee 37996-1200, USA }
\author{G.F. Chen}
\affiliation{Institute of Physics, Chinese Academy of Sciences, Beijing 100080, China}
\author{Chen Fang}
\affiliation{Department of Physics, Purdue University, West Lafayette, IN 47907, USA}
\author{T.G. Perring}
\affiliation{ISIS Facility, STFC Rutherford Appleton Laboratory, Didcot, Oxfordshire
OX11 0QX, UK}
\affiliation{Department of Physics and Astronomy, University College London, London WC1E 6BT, UK}
\author{D.L. Abernathy}
\affiliation{Oak Ridge National Laboratory, Oak Ridge, Tennessee 37831}
\author{A.D. Christianson}
\affiliation{Oak Ridge National Laboratory, Oak Ridge, Tennessee 37831}
\author{Takeshi Egami}
\affiliation{The University of Tennessee, Knoxville,
Tennessee 37996-1200, USA }
\affiliation{Oak Ridge National Laboratory, Oak Ridge, Tennessee 37831}
\author{Nanlin Wang}
\affiliation{Institute of Physics, Chinese Academy of Sciences, Beijing 100080, China}
\author{Jiangping Hu}
\affiliation{Department of Physics, Purdue University, West Lafayette, Indiana 47907, USA}
\affiliation{Institute of Physics, Chinese Academy of Sciences, Beijing 100080, China}
\author{Pengcheng Dai}
\email{daip@ornl.gov}
\affiliation{The University of Tennessee, Knoxville,
Tennessee 37996-1200, USA }
\affiliation{Oak Ridge National Laboratory, Oak Ridge, Tennessee 37831, USA}
\affiliation{Institute of Physics, Chinese Academy of Sciences, Beijing 100080, China}

\begin{abstract}
We use neutron scattering to show that spin waves in the iron chalcogenide Fe$_{1.05}$Te display novel dispersion clearly different from both the first principle density functional calculations and recent observations in the related iron pnictide CaFe$_2$As$_2$.  By fitting to a Heisenberg Hamiltonian, we find that although the nearest-neighbor exchange couplings in the two systems are quite different, their next-nearest-neighbor (nnn) couplings are similar.  This suggests that superconductivity in the pnictides and chalcogenides share a common magnetic origin that is intimately associated with the nnn magnetic coupling between the irons.
\end{abstract}


\maketitle

\begin{figure}
\begin{center}
\includegraphics[width=0.75\linewidth]{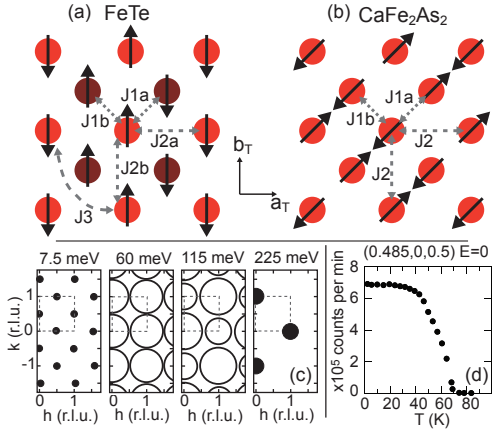}
\end{center}\caption{(a) Schematic of in-plane Fe spins displaying magnetic order in Fe$_{1+y}$Te with small $y$ \cite{Bao2009,Li2009}, and showing definition used for exchange energies. (b) Schematic of in-plane magnetic order in CaFe$_2$As$_2$ \cite{Goldman2008} with exchange energy definitions. (c) Schematic showing wave vector dependence of intensity at various energies (for raw data see Fig. \ref{Fig:fig2}). Dashed line shows one BZ. (d) Temperature dependence of elastic scattering at magnetic Bragg peak for the Fe$_{1.05}$Te sample.}
\label{Fig:fig1}
\end{figure}

All parent compounds of cuprate superconductors are antiferromagnetic (AF) Mott insulators characterized by the same local moment Heisenberg Hamiltonian \cite{Coldea2001}. For this reason,
it is believed that magnetism
is important for the high-$T_c$ superconductivity \cite{Lee2006}. The iron-based superconductors \cite{Kamihara2008,Johnston2010} share many features in common with the cuprates, which leads many to conjecture that the magnetism present in these compounds is vital
for the presence of superconductivity. The iron-based superconductors can be divided into two chemical classes, the iron pnictides such as CaFe$_2$As$_2$ and iron chalcogenides Fe$_{1+y}$Te. Many properties of the pnictides and chalcogenides are similar, including similar band-structure \cite{Subedi2008} and magnetic excitations in the superconducting compositions \cite{Christianson2008,Lumsden2009,Chi2009,Inosov2010,Mook2010,Qiu2010,Liu2010}. Furthermore, the magnetism in the pnictide parent CaFe$_2$As$_2$ [Fig. \ref{Fig:fig1}(b)] is consistent with first principle density functional calculations \cite{Han2009b}. However, the parent compound \cite{Bao2009,Li2009} of the iron chalcogenides, Fe$_{1+y}$Te, possesses a different AF order [Fig. \ref{Fig:fig1}(a)]. Therefore,
it is important to determine if magnetism in these two systems can be described by a similar Hamiltonian. If the magnetic description between systems is entirely dissimilar, then it presents a serious challenge to many theories \cite{Mazin2009,Kuroki2008,Seo2008,Wang2009} where superconductivity has a magnetic origin.

By studying the spin-waves in Fe$_{1.05}$Te, we compare the magnetic couplings within the pnictide and chalcogenide systems. We show that although the nearest neighbor (nn) couplings in the two systems are very different, the effective next nearest couplings (nnn) $J_2$ are very similar. 
While our results are consistent with the theoretical idea that $J_2$ is important for superconductivity \cite{Seo2008}, 
 the isotropic $J_2$ we find in Fe$_{1.05}$Te is very different from the anisotropic $J_2$ yielded from density functional calculations \cite{Han2009}. Our results suggest that while the nn coupling may change, it is the nnn coupling that persists between different iron superconductors.

We have used time-of-flight inelastic neutron spectroscopy to determine the dispersion of spin-wave excitations in Fe$_{1.05}$Te (with AF ordering temperature $T_N =$ 68~K, see Fig. \ref{Fig:fig1}(d) and ref. \onlinecite{Chen2009}), the $x = 0$ (non-superconducting) member of the isovalently substituted Fe$_{1+y}$Te$_{1-x}$Se$_x$ iron chalcogenide superconductors \cite{Hsu2008,Fang2008}.  By measuring spin-wave excitations in Fe$_{1.05}$Te throughout the Brillouin zone (BZ), we have used a Heisenberg Hamiltonian to determine the effective exchange couplings of the system.  Our neutron scattering experiments were carried out on the HB-1 triple-axis spectrometer at High-Flux-Isotope-Reactor and on the ARCS chopper spectrometer at Spallation-Neutron-Source, Oak Ridge National Laboratory, USA.  We also used MAPS chopper spectrometer at ISIS, Rutherford-Appleton Laboratory, UK.  For the experiment, we have co-aligned 6~grams of single crystals of Fe$_{1.05}$Te. All data were collected at around 10~K ($\ll T_N$) with incident neutron energies $E_i =$ 55, 90, 180, 350, 500 and 580~meV with the $c$-axis aligned along the incident beam direction.  Since the spin-wave excitations have weak $c$-axis coupling, we integrate the excitations along the $c$-axis direction, and focus on spin waves in the $(h,k)$ plane.  

For Fe$_{1+y}$Te with modest excess iron content $y$, the magnetic structure is shown in Fig. \ref{Fig:fig1}(a) \cite{Bao2009,Li2009}, which can be viewed as two AF sub-lattices as shown by darker and lighter colored atoms.  We define the nn ($J_{1a}$, $J_{1b}$), the nnn ($J_{2a}$, $J_{2b}$), and the next-next-nearest neighbor ($J_3$) exchange interactions as shown in Fig. \ref{Fig:fig1}(a) \cite{Han2009}. The nn magnetic exchange couplings ($J_{1a}$, $J_{1b}$) are defined similarly to those of iron pnictides [Fig. \ref{Fig:fig1}(b)].  However, the nnn couplings ($J_{2a}$, $J_{2b}$) in chalcogenides are directionally dependent as shown in Fig. \ref{Fig:fig1}(a).

Our Fe$_{1.05}$Te samples
were grown using Bridgman technique as described before \cite{Chen2009}.  Fe$_{1+y}$Te$_{1-x}$Se$_x$ is tetragonal at high temperature and becomes orthorhombic or monoclinic (depending on $x$, \cite{Hsu2008,Fang2008,Bao2009,Li2009}) below $T_N$.  The $ab$-plane lattice parameters for the various phases remain very similar, and on cooling into the low symmetry phase the sample becomes twinned.  We therefore measure the wave vector in tetragonal $(h,k,l)$ reciprocal lattice units,
with in-plane lattice parameters $a=b=$3.80~\AA, and the out-of-plane $c = 6.23$~\AA.  In this notation, magnetic order in powder Fe$_{1+y}$Te has been found at $(0.5,0,0.5)$ for small $y$, and increasing $y$ will lead to incommensurate magnetic order \cite{Bao2009,Li2009}.  In the present single crystalline samples, the magnetic order was found to be centered very close to the commensurate position at $(0.485,0,0.5)$~r.l.u and $y=$0.05 was measured with inductively coupled plasma analysis \cite{Chen2009}.  However, we also observed a weaker magnetic peak at $(0.37,0,0.5)$~r.l.u attributed to a small portion of the sample with slightly different $y$.  Figure \ref{Fig:fig1}(d) shows the temperature dependence of the magnetic Bragg intensity at $\mathbf{Q} = (0.485,0,0.5)$~r.l.u confirming T$_N=$ 68~K.

\begin{figure}
\begin{center}
\includegraphics[width=0.99\linewidth]{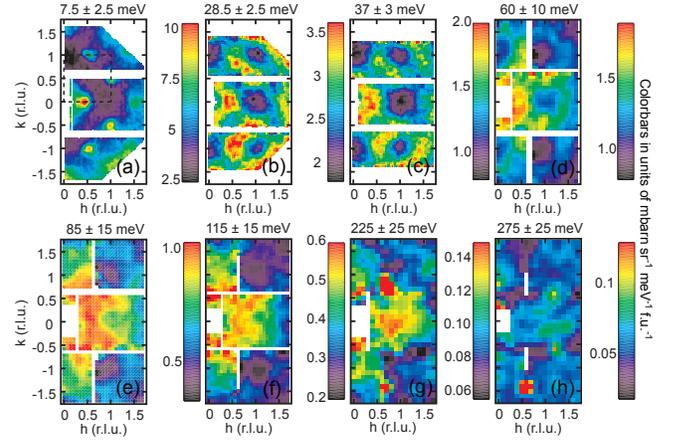}
\end{center}\caption{Constant energy slices of the spin-waves as a function of increasing energy at 10~K for Fe$_{1.05}$Te.  All data are normalized to absolute units with a vanadium standard.  (a)-(c) collected with incident neutron energy $E_i =$ 90~meV on ARCS, (d)--(f) $E_i =$ 350~meV on MAPS, (g)--(h) $E_i =$ 500~meV on MAPS. The dashed line in (a) shows a crystallographic BZ.}
\label{Fig:fig2}
\end{figure}

The magnetic excitations probed by neutron scattering in our Fe$_{1.05}$Te sample are summarized by representative constant energy slices in Fig. \ref{Fig:fig2}.  The data have been normalized to a vanadium standard and plotted in absolute units, without correction for the magnetic form factor, causing the signal intensity to decrease with increased $Q$.  Each $E_i$ probes a different out-of-plane wave vector for each energy transfer, and it was found that data from different $E_i$'s were consistent, implying little $L$-dependence of the data over the energy range probed.

Spin waves in most materials tend to display a magnetic response centered on the magnetic Bragg position up to the highest energies, with successively larger rings with increased energy. However, we discuss below how the center of the excitations switch from the $(0.5,0)$ low energy position to integer positions at higher energy, which we interpret as the outcome of the interaction of competing ferromagnetic and AF exchange energies.

\begin{figure}
\begin{center}
\includegraphics[width=0.9\linewidth]{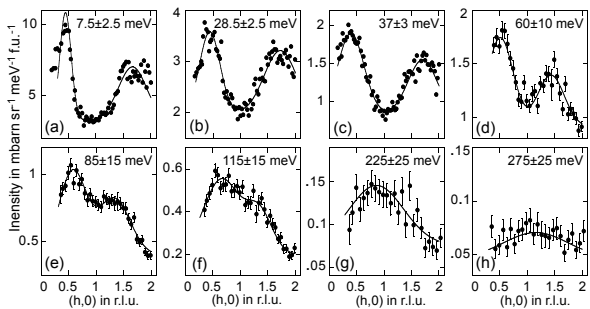}
\end{center}\caption{Constant energy cuts along the $(h,0)$ trajectory, each from a slice in Fig. \ref{Fig:fig2}. Solid lines are fits to Gaussians.
}
\label{Fig:fig3}
\end{figure}

At our lowest energy, 7.5~meV [Fig. \ref{Fig:fig2}(a)], magnetic excitations emerge from the AF Bragg position $(0.5,0)$ and other half-integer reciprocal lattice vectors [in an untwinned sample, magnetic peaks would not appear at $(0,0.5)$, but twinning leads to an equal intensity domain rotated by 90$^\circ$ in-plane].  As the energy is increased, the response spreads out in $Q$ as expected for spin-waves [Figs. \ref{Fig:fig2}(b) and \ref{Fig:fig2}(c)].  As the energy is raised to around 60~meV [Fig. \ref{Fig:fig2}(d)], there are no longer peaks at half-integer positions, but instead there are rings of radii $\sim$0.5~r.l.u which are centered on integer reciprocal lattice points. These rings are even clearer when the data are corrected for the magnetic form factor drop-off at high wave vector (see supplementary material).
As energy is increased, the radii of rings around $(1,1)$ expand and those around $(1,0)$ contract [Figs. \ref{Fig:fig2}(e)--\ref{Fig:fig2}(f)].  Even at 115~meV a ring can be seen around $(1,0)$, which by 225~meV contracts into a peak at $(1,0)$ [Fig. \ref{Fig:fig2}(g)] before the disappearance of all intensity at higher energies [Fig. 2(h)].  Corresponding cuts along the $(h,0)$ trajectory are shown in Fig. \ref{Fig:fig3}.  A schematic of the dispersion of the magnetic response is shown in Fig. \ref{Fig:fig1}(c).  The data above 100~meV in Fe$_{1.05}$Te have similarities to the highest energy spin excitations observed in FeTe$_{1-x}$Se$_x$ with $x =$ 0.27, 0.49 \cite{Lumsden2010}.

In order to extract effective exchange energies, we fit spin-wave data using a Heisenberg Hamiltonian (see supplementary material for the model Hamiltonian) with commensurate $(0.5,0,0.5)$ AF \cite{Fang2009}.  In order to yield this commensurate AF, there are constraints on the bounds of each of the magnetic exchange energies \cite{Fang2009}.  Because of the twinned nature of the sample, the model used is the sum of two equal sized domains rotated by 90$^\circ$.

\begin{figure}
\begin{center}
\includegraphics[width=0.70\linewidth]{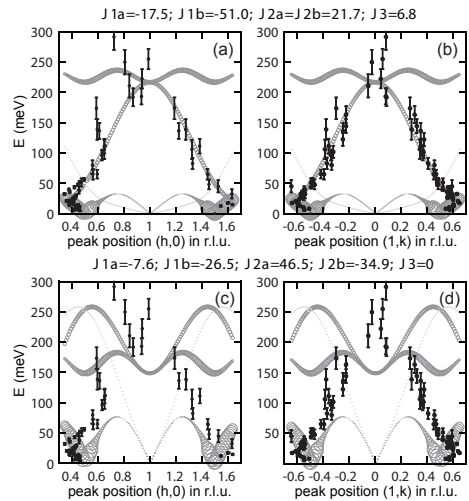}
\end{center}\caption{(a)--(b) Solid black markers are dispersion data found from fitting Gaussians to form factor corrected data at many energies for the $(h,0)$ and $(1,k)$ directions respectively.  Gray open circles (with radius indicating intensity) show best fit dispersion curves with fitting parameters given in the main text. (c)--(d) Data as in (a)--(b), but with dispersion curves simulated using exchange constants predicted by density functional calculations, which clearly do not agree with the data.}
\label{Fig:fig4}
\end{figure}

To determine the dispersion curves for spin waves, the slices in Fig. \ref{Fig:fig2} were cut along the $(h,0)$ and $(1,k)$ directions. 
By fitting Gaussians to many $(h,0)$ cuts of different energies like those in Fig. \ref{Fig:fig3}, we obtain  the dispersion plot in Fig. \ref{Fig:fig4}(a) using the fitted peak positions.  Similarly, $(1,k)$ cuts were fitted to create Fig. \ref{Fig:fig4}(b).  These two dispersion plots were simultaneously fitted to the dispersion of the model \cite{Fang2009}, yielding the fit displayed in Figs. \ref{Fig:fig4}(a)--(b). Similar conclusions about the dispersion could be reached by viewing the data in terms of constant-$Q$ cuts instead of cuts at constant energy, but this was not found to be as effective for quantitative analysis (see supplementary material).
In Fig. \ref{Fig:fig4}, the intensity of the excitations of the model is proportional to the radius of the marker (which is saturated at the lowest energies to maintain figure clarity), to highlight the bands with negligible intensity (also see the supplementary material for a zoom into the low energy part of the plots).
The presence of almost non-dispersive bands around 250~meV is not clear in the $Q$-cuts, possibly because of averaging-out in $Q$ as the bandwidth is comparable to the instrument resolution (along with poorer statistics at high energies). It is also not clear if these bands can be seen in constant-$Q$ analysis (see supplementary material).

In the fit lines displayed in Fig. \ref{Fig:fig4}(a)--(b), $J_{2b}$ was fixed equal to $J_{2a}$, after it was found that these two parameters had very similar values when allowed to vary (see supplementary material for fit with $J_{2b}$ not fixed to $J_{2a}$).  This four parameter fit leads to exchange energies of $J_{1a} =$ -17.5 $\pm$ 5.7, $J_{1b} =$ -51.0 $\pm$ 3.4, $J_2 = J_{2a} = J_{2b} =$ 21.7 $\pm$ 3.5, $J_3 =$ 6.8 $\pm$ 2.8~meV (assuming $S$ = 1) and fits the dispersion in these directions well.  By further fixing $J_3 =$ 0, the model can successfully fit the data up to $\sim$ 100~meV, but the maximum band energy, $\omega_{\rm max}$, is underestimated by around 50~meV (see supplementary material for fits where $J_3$ is fixed to zero).

\begin{figure}
\begin{center}
\includegraphics[width=0.95\linewidth]{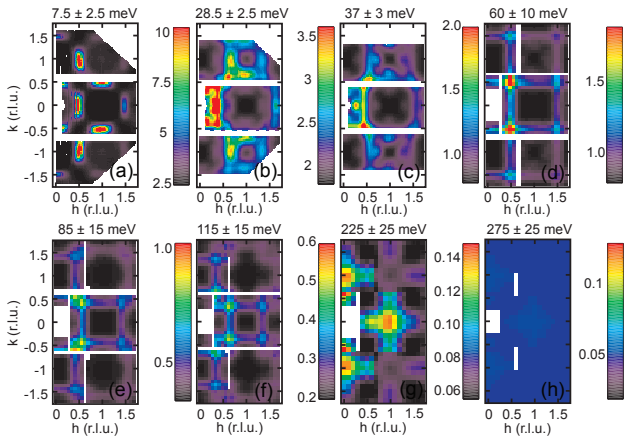}
\end{center}\caption{Resolution convolved simulation (using Tobyfit \cite{TobyFit}) of the Heisenberg model using the best fit parameters in the text plus an out of plane coupling of $J_z =$ 1~meV.  Each slice corresponds to a slice in Fig. \ref{Fig:fig2}. The model has been given a line-width of 10~meV before resolution convolution, though adjusting the line-width does not make a substantial difference.  All slices are on the same intensity color scale as Fig. \ref{Fig:fig2}, with an overall intensity scale that was chosen so that intermediate simulation slices had a similar intensity to the intermediate raw data slices.}
\label{Fig:fig5}
\end{figure}

Using the fit parameters listed above, we show in Fig. \ref{Fig:fig5} constant energy slices calculated from the resolution-convolved model.  Here we have also considered the out-of-plane ($c$-axis) exchange coupling $J_z$ and found that $J_z =$ 1~meV best fits the spin-wave intensities, although the simulation slices otherwise do not change significantly with $J_z$. The overall features of the model fit are: (i) below $\sim$30~meV, intensity is located around $(0.5,0)$; (ii) at intermediate energy there are rings around $(1,1)$ that grow with increasing energy; (iii) above $\sim$150~meV the intensity ends in a peak at $(1,0)$.  The data are consistent with the model, though the intermediate energy features are more grid-like than the more rounded data.

Our fits and simulations show highly anisotropic in-plane nn exchange couplings with $|J_{1b}| \gg |J_{1a}|$, and a nnn exchange that is AF (energy $\sim$20 meV) and isotropic $J_2 = J_{2a} \approx J_{2b}$.  The $\omega_{\rm max}$ observed is between 200--250~meV.  Comparing our results to similar high energy measurements of CaFe$_2$As$_2$ \cite{Zhao2009}, which has $J_{1a} =$ 50$\pm$ 10, $J_{1b} =$ -5.7$\pm$ 5, $J_2 =$ 19$\pm$ 3~meV and $\omega_{\rm max}\approx$200~meV, it is clear that the $\omega_{\rm max}$ and values of $J_2$ are similar, as well as the presence of anisotropy in $J_1$ in both cases plus no anisotropy in $J_2$ in either case.  However, the dominating $J_1$ exchange constants are -50~meV ($J_{1b}$) and +50~meV ($J_{1a}$) for Fe$_{1.05}$Te and CaFe$_2$As$_2$, respectively.

Our results shed new light on the nature of the magnetic state in the iron chalcogenides and its relationship to superconductivity.  The isotropic $J_2$ suggests that this nnn exchange coupling originates from the super-exchange mechanism, and is insensitive to the lattice distortion and variation in the $d$-orbital components.  Theoretically, it has been shown that the nnn \cite{Seo2008} magnetic coupling can cause an s$^{\pm}$-wave pairing that induces a neutron spin resonance at wave vector $(0.5,0.5)$ \cite{Korshunov2008,Maier2009}.  Similar isotropic AF $J_2$ values in iron-pnictides and iron chalcogenides therefore naturally explain the experimentally observed neutron spin resonance within both classes of iron-based superconductors \cite{Christianson2008,Lumsden2009,Chi2009,Inosov2010,Mook2010,Qiu2010,Liu2010}.  First principles density functional calculations \cite{Han2009} on Fe$_{1.068}$Te predict highly anisotropic nnn exchange interactions which are not consistent with our data [see Figs. 4(c) and 4(d) for dispersion and the supplementary material for simulation slices], perhaps due to the complex nature of the orbital ordering \cite{Lv2010,Kruger2009} or itinerant magnetism \cite{Moon2010} in this material.

In summary, we have shown that spin-wave excitations in the iron chalcogenide Fe$_{1.05}$Te can be modeled by a Heisenberg Hamiltonian with anisotropic (dominantly) ferromagnetic nn and isotropic AF nnn exchange couplings.  While the nn couplings for Fe$_{1.05}$Te and CaFe$_2$As$_2$ \cite{Zhao2009} are different, we find that the AF nnn exchange couplings in these two classes of materials are not only similar in magnitude but also directionally independent, even though they have different AF and crystal structures \cite{Bao2009,Li2009,Goldman2008}.  Our findings suggest that superconductivity in both classes of iron-based superconductors shares a common magnetic origin that is intimately associated with the AF nnn exchange couplings \cite{Seo2008}.  

This work is supported in part by the US DOE, BES, through DOE DE-FG02-05ER46202 and by the US DOE, Division of Scientific User Facilities.  The work at the IOP is supported by the CAS.  OJL and TE were supported by the DOE, BES, EPSCoR Grant DE-FG02-08ER46528.

\appendix

\makeatletter \renewcommand{\thefigure}{S\@arabic\c@figure} \renewcommand{\thetable}{S\@arabic\c@table} \makeatother
\setcounter{figure}{0} 


\section{Data corrected for Fe$^{2+}$ magnetic form factor}

\begin{figure}
\begin{center}
\includegraphics[width=0.99\linewidth]{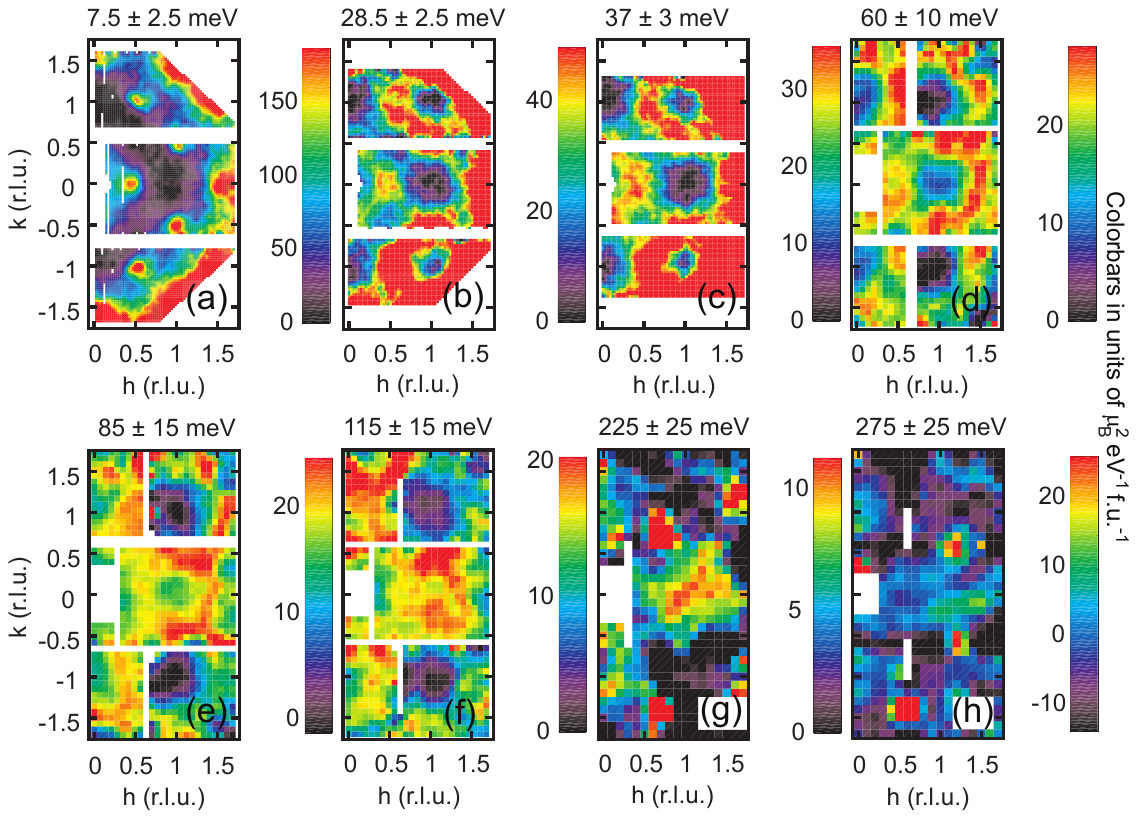}
\end{center}\caption{Constant energy slices of data (from Fig. 2) converted into units of magnetic response $\chi^{''}(\mathbf{q},\omega)$ (after subtracting constant backgrounds).  After the form factor correction, the magnetic signal no longer decreases in intensity at high wave vector transfer, and so the magnetic pattern is somewhat clearer than in the raw data.  However, due to a slight increase in background at large wave vectors (backgrounds tend to gradually, and reasonably isotropically, increase with higher $Q$ due to multiphonon scattering), the slices [notably (a)-(c)] show excessive intensity at the highest wave vectors.}
\label{Fig:figS1}
\end{figure}

To see more clearly the evolution of spin waves in Fe$_{1.05}$Te in Fig. 2 of the main text, we plot in Fig. S1 constant energy slices of the data in Fig. 2 converted into units of magnetic response,  $\chi^{''}(\mathbf{q},\omega)$.  This is related to the measured cross-section by
\begin{equation}\label{Eq:crosssection}
\frac{k_i}{k_f} \frac{d^2 \sigma}{d\Omega dE} = \frac{2(\gamma r_e)^2}{\pi g^2 \mu_B^2} |F(\mathbf{Q}|^2 \frac{\chi^{''}(\mathbf{q},\omega)}{1-\exp(-\hbar\omega / k T)}
\end{equation}
where $F(\mathbf{Q})$ is the Fe$^{2+}$ magnetic form factor, $(\gamma r_e)^2$=0.2905 b sr$^{-1}$, $k_i$ and $k_f$ are initial and final neutron wave vector and the $g$-factor was assumed to be 2.  The exponential term is from the Bose factor, which can be neglected at the energies of interest, as all measurements were taken at 10~K.  The left hand side of this equation is the raw data as plotted in Fig. 2.

\section{Heisenberg Hamiltonian}

For a given set of exchange energies ($J_{1a}$, $J_{1b}$, $J_{2a}$, $J_{2b}$, $J_3$, $J_z$) as shown in Fig. 1(a), the dispersion of magnetic excitations from the Heisenberg Hamiltonian with magnetic order at $(0.5,0,0.5)$ can be found by diagonalizing the Hamiltonian $H$ \cite{Fang2009} for every Miller index $(h, k, l)$ value: 
\begin{equation}\label{Eq:Ham}
H=
\left( \begin{array}{cccc}
A & J_{1b}D & -B & -J_{1a}C\\
J_{1b}C& A &-J_{1a}D &-B\\
B &J_{1a}C &-A &-J_{1b}D\\
J_{1a}D& B &-J_{1b}C& -A\\
\end{array} \right)
\end{equation}
where
\begin{eqnarray}\label{Eq:Ham2}
& A=2 J_{2b} \cos(k_x-k_y) + 2 (J_{1a} + J_{2a}-J_{1b}-J_{2b}+2J_3+J_z)\nonumber \\
& B=2 J_{2a} \cos(k_x+k_y) + 2J_3 [\cos(2k_x) + \cos(2k_y)]+2J_z \cos k_z]\nonumber \\
& C=e^{i k_x} + e^{i k_y}\nonumber \\
& D=e^{-i k_x} + e^{-i k_y}\nonumber
\end{eqnarray}
with
\begin{eqnarray}\label{Eq:Ham3}
k_x=\pi (h + k)\nonumber \\
k_y=\pi (h-k)\nonumber \\
k_z=2\pi l
\end{eqnarray}

For a twinned sample, the full dispersion must be calculated at both $(h,k,l)$ and $(-k,h,l)$. The two Fe atoms per unit cell leads to two bands per wavevector, which after twinning leads to a total of four bands. However, at least one of these bands has zero intensity at the wavevector positions we compute the dispersion.

\section{The isotropy of $J_2$ and the effect of $J_3$ on the spin-wave fits}

As described in the paper, the dispersion data in the $(h,0)$ and $(1,k)$ directions were obtained from fitting wave vector peak positions of many constant energy cuts.  These two dispersion plots could then be simultaneously fitted to the commensurate phase of the Heisenberg model to yield the exchange energies $J_{1a}$, $J_{1b}$, $J_{2a}$, $J_{2b}$, and $J_3$.  The parameters in the fits were constrained to bounds such that magnetic order occurs at the $(0.5,0,0.5)$ point, a requirement of the magnetic model (see the sections below and Ref. \onlinecite{Fang2009} for more details).

\begin{figure}
\begin{center}
\includegraphics[width=0.7\linewidth]{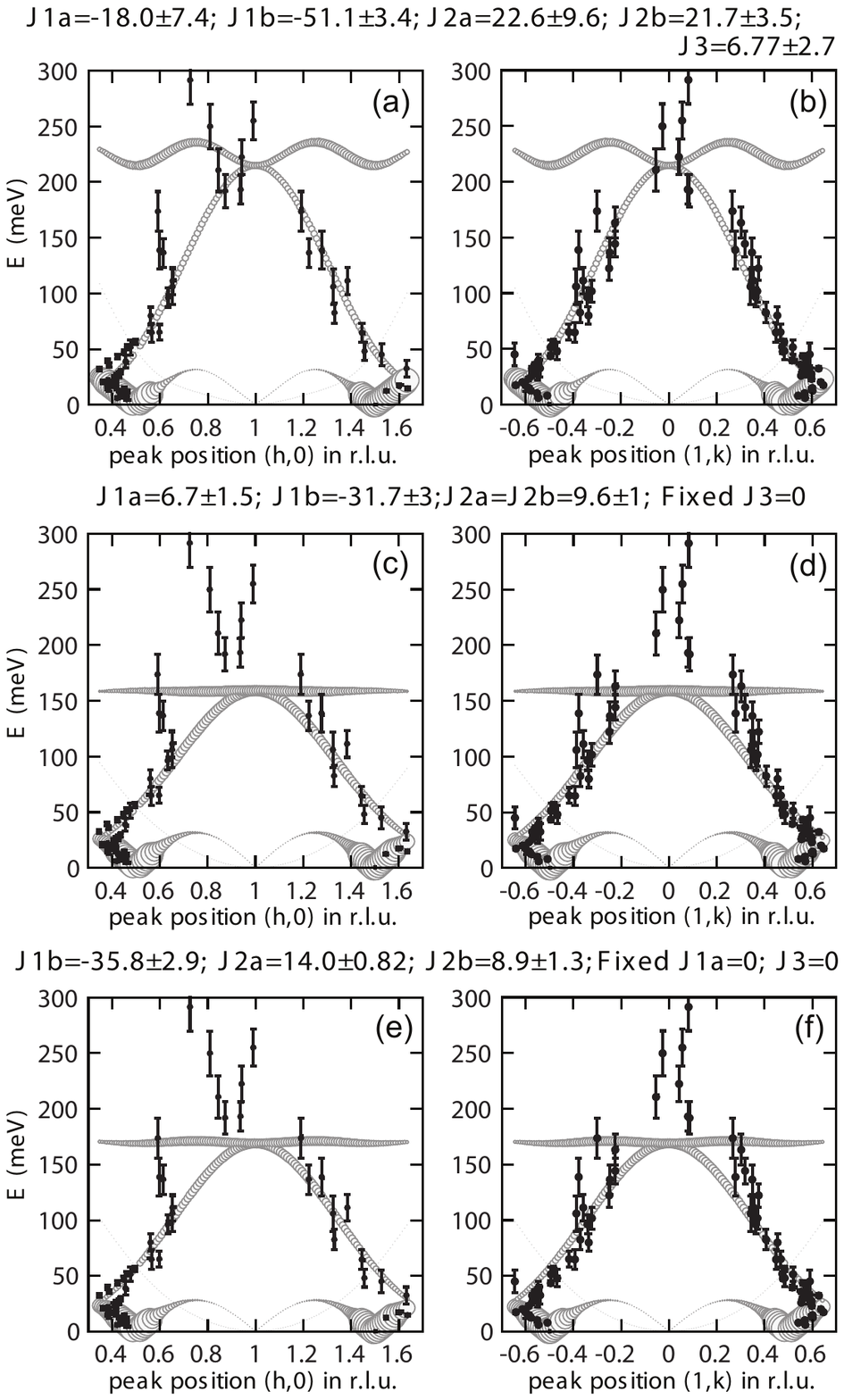}
\end{center}\caption{Dispersion extracted from data (solid symbols) and fits of the model dispersion (gray open circles with radius indicating intensity). Left panels show dispersion in $(h,0)$ direction, and right panels show $(1,k)$ dispersion. (a)--(b) Fit performed with all parameters free. (c)--(d) Fit with fixed $J_3 = 0$ and $J_{2a} = J_{2b}$. (e)--(f) Fit with fixed $J_3 = 0$ and $J_{1a} = 0$.}
\label{Fig:figS2}
\end{figure}

\begin{figure}
\begin{center}
\includegraphics[width=0.8\linewidth]{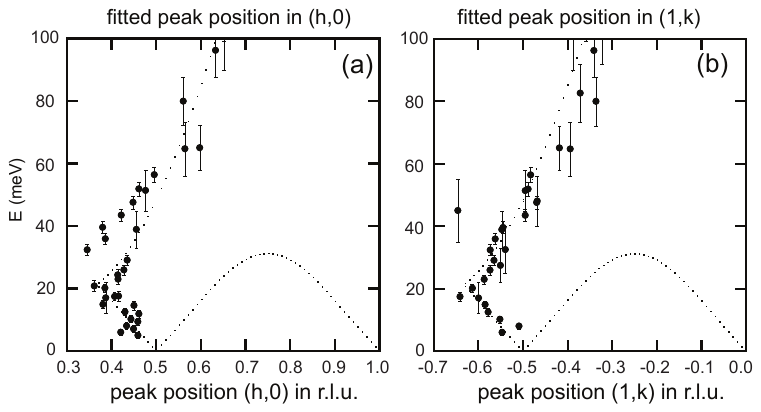}
\end{center}\caption{Dispersion extracted from data (solid symbols) and model from best fit parameters (dotted line), zoomed in to emphasize low energy region of Fig. 4(a) and Fig. 4(b) respectively. Though for clarity the intensity of the model dispersion is not shown in this plot, it can be seen from the marker sizes in Fig. 4(a)--(b) that the intensity of the branches to the right of 0.5 (and -0.5 respectively) in these figures would die away very quickly with increasing energy. Furthermore, at the lowest energies the two branches cannot be resolved. The outcome is that the branch to the right of the two figures are not observed in the data.}
\label{Fig:figS2p5}
\end{figure}

\begin{figure}
\begin{center}
\includegraphics[width=0.6\linewidth]{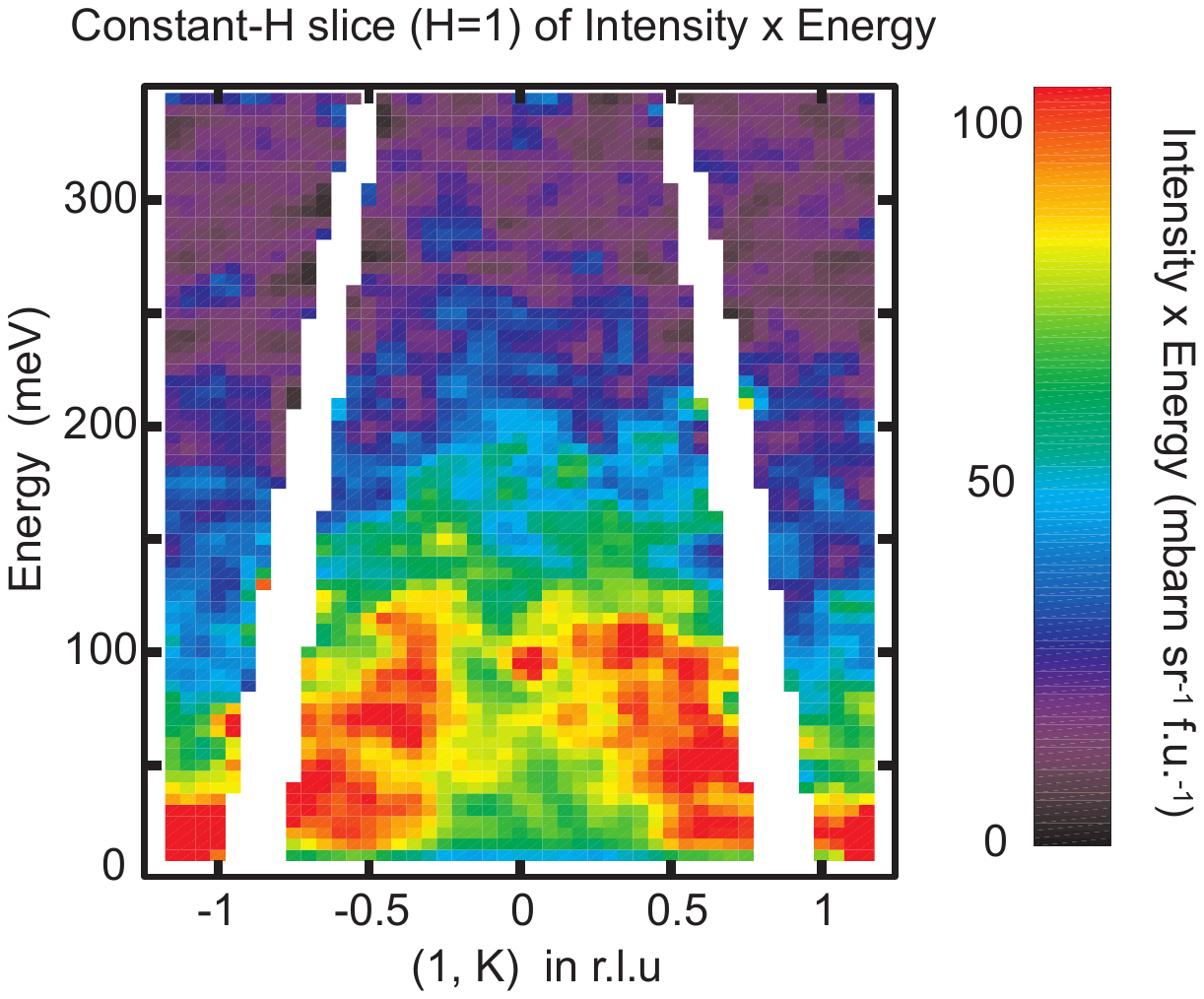}
\end{center}\caption{Dispersion as shown by a slice of the raw data (multiplied by energy transfer to enhance high energies) with a fixed $h=1$, from the MAPS $E_i=500$~meV run. The dispersion into $(1,0)$ around 200~meV is clearly visible.}
\label{Fig:figS2p6}
\end{figure}

To start with, the dispersion was fitted by varying all $J_{1a}$, $J_{1b}$, $J_{2a}$, $J_{2b}$, and $J_3$ parameters.  This yielded the fits in Fig. S2(a)--(b) which gave $J_{1a} = -18.0\pm7.4$; $J_{1b} = -51.1\pm3.4$; $J_{2a} = 22.6\pm9.6$; $J_{2b} = 21.7\pm3.5$; $J_3 = 6.77\pm2.7$~meV.  This fit was robust, and as it was clear that $J_{2a}$ and $J_{2b}$ were very close, they were subsequently fixed to be equal and refitted.  This fit ($J_{1a} = -17.5\pm5.7$; $J_{1b} = -51.0\pm3.4$; $J_2 = J_{2a} = J_{2b} = 21.7\pm3.5$; $J_3 = 6.8\pm2.8$~meV) [shown in Fig. 4(a)--(b)] clearly yields parameters almost the same as the original fit parameters as expected, and is quoted as the main result in the paper. The low energy region of Fig. 4(a)--(b) is shown scaled up in Fig. \ref{Fig:figS2p5}.

\begin{figure}
\begin{center}
\includegraphics[width=0.99\linewidth]{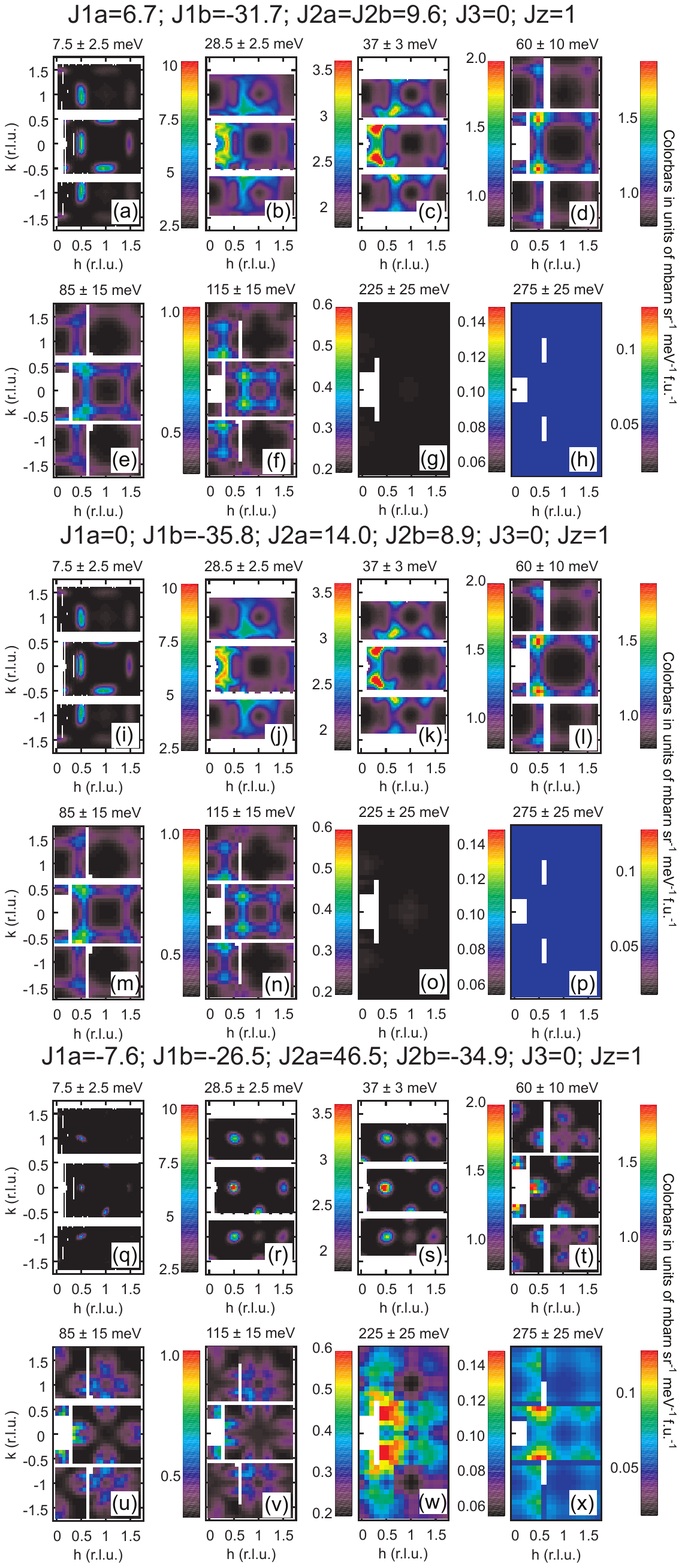}
\end{center}\caption{Instrument resolution convolved simulation of model with various exchange energy parameters. (a)--(h) fit with $J_3$ fixed to zero (and $J_{2a}=J_{2b}$), with dispersion shown in Fig. S2(c)--(d). (i)--(p) fit with $J_3$ and $J_{1a}$ fixed to zero, with dispersion shown in Fig. S2(e)--(f). (q)--(x) parameters from the Han et al. \cite{Han2009} prediction, the dispersion of which was shown in Fig. 4(c)--(d).  For comparison with the simulation with best fit parameters shown in Fig. 5.  $J_z$ does not vary the dispersion significantly (see main text of paper) and was therefore fixed for all plots to $J_z = 1$~meV, the value chosen in Fig. 5. The intensity of the model in (q)--(x) was magnified by a factor of 1.5 in order to show the pattern more clearly.}
\label{Fig:figS_simAll}
\end{figure}

Before discussing the necessity of the next-next-nearest neighbor term \cite{Ma2009}, $J_3$, we will first briefly compare our dispersion with the theoretical exchange energies predicted by Han et al. \cite{Han2009}.  Figures S2(g)--(h) shows our experimental data compared with the dispersion calculated for their theoretical values, and it is clear that the data and model are very dissimilar.  The model cannot be reconciled with the data because in this theory $J_{2b} = -J_{2a}$, whereas our data appears to be best described with $J_{2a} = J_{2b}$.

Following our successful fit with $J_{2a} = J_{2b}$, it was important to check whether the small next-next-nearest neighbor parameter $J_3$ could be fixed to zero in order to further decrease the number of parameters in the fit, and see if this extra small exchange interaction is actually necessary or not.  The obvious starting place is to calculate the dispersion when using the above fit parameters but with $J_3$ set to zero.  However, the model dispersion cannot be calculated in this case because this shifts the parameters out of the range allowed by $(0.5,0,0.5)$ magnetism \cite{Fang2009}.  So instead, we performed another fit, whilst fixing $J_3$ to zero.  In Figures S2(c)--(d), this fit can be seen, yielding $J_{1a} = 6.7\pm1.5$; $J_{1b} = -31.7\pm3$; $J_2 = J_{2a} = J_{2b} = 9.6\pm1.0$~meV (and $J_3 = 0$). This fit describes the data below 100~meV well, but is not as successful at the high energies, underestimating the maximum energy of the band by around 50~meV.  We interpret this fit as compensating the lack of an AF $J_3$ term by instead causing the $J_{1a}$ term to be AF.  The compromise is that $J_{1a}$ cannot be too large or the high energy data cannot be described.

\begin{figure*}
\begin{center}
\includegraphics[width=0.6\linewidth]{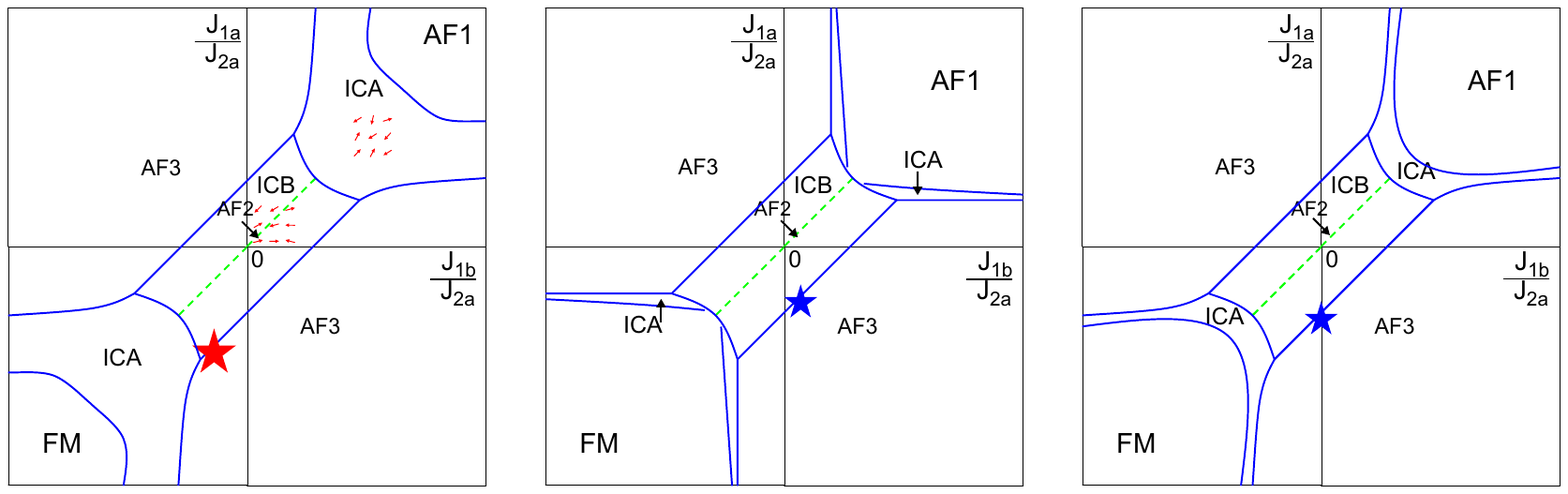}
\end{center}
\caption{Slices of the schematic phase diagram of the $J_1-J_2-J_3$ model. The phases are defined in ref. \onlinecite{Fang2009}, among which the AF3 is the commensurate SDW phase observed in the experiment. ICA and ICB are incommensurate magnetic phases, and FM is a ferromagnetic phase. The red star in (a) shows the position in phase space of the best fit parameters shown in Fig. S2(a)--(b). The stars in (b) and (c) show the position in phase space of the other fits in Fig. S2(c)--(d) and Fig. S2(e)--(f) respectively. Note that the lower left corner gives the ferromagnetic phase, obtained from the $(0.5,0.5)$ order (AF1) phase by a sublattice symmetry $\hat{S}$ defined above.}
\end{figure*}

One may finally ask whether the data could be fitted better if $J_3$ was fixed to zero and $J_{2a}$ and $J_{2b}$ could vary independently.  In this case $J_{1a}$ was found to stay small and so was fixed to zero for ease of fitting (results do not vary significantly with a fitted $J_{1a}$ term). The fit is shown in Fig. S2(e)--(f) (with parameters $J_{1b} = -35.8\pm2.9$; $J_{2a} = 14.0\pm0.82$; $J_{2b} = 8.9\pm1.3$, and $J_{1a} = 0$; $J_3 = 0$~meV).  In this fit it can be seen that (i) $J_{2a}$ and $J_{2b}$ are still quite similar, and (ii) the fit is also not as good as the original ($J_3 > 0$) fit, as it again underestimates the maximum energy.

We conclude that the data is best fit with an isotropic $J_2$ ($= J_{2a} = J_{2b}$).  The addition of a $J_3$ exchange interaction also seems to be important to describe the highest energies whilst keeping the parameters in the regime with $(0.5,0,0.5)$ magnetic order.  For further insight into the similarity and dissimilarities between these different fits and models, we show instrument resolution convolved simulations with $J_3$ fixed to zero [see Fig. S5(a)--(h) for simulation with parameters from the fit in Fig. S2(c)--(d), and see Fig. S5(i)--(p) for simulation with parameters from the fit in Fig. S2(e)--(f)].  These alternative parameters show similar features to our data [compare with simulation of original ($J3 > 0$) fit in Fig. 5], though do not extend high enough in energy.  In Fig. S5(q)--(x) we have also included a simulation with the Han et al. \cite{Han2009}, theoretical parameters [see Fig. S2(g)--(h)] which have $J_{2b} = -J_{2a}$.  It is clear that these parameters do not describe the data well.

\section{Viewing the data from constant-$Q$ perspective}

Performing energy cuts instead of $(h,0)$ and $(1,k)$ cuts does not show the dispersion as clearly as $Q$-cuts because of the decrease in signal as energy increases. However the overall dispersion is clear when plotted as intensity multiplied by energy in constant-$Q$ slices, see Fig. \ref{Fig:figS2p6}. It is still unclear in this slice whether the 250~meV almost non-dispersive band exists or not.

\section{The extended phase diagram}

In this section, we extend the phase diagram of the $J_1-J_2-J_3$ model (see ref. \onlinecite{Fang2009} and Hamiltonian described in a section above) applied to FeTe$_{1-x}$Se$_{x}$ system to the parameter region where $J_{1a,1b}$ can be negative (providing $J_{2a}\geq J_{2b}>0$). First, it is only trivial to observe that the original model has the following two properties:
\bea U(\hat{S})^\dag{H(J_{1a},J_{1b},J_{2a},J_{2b},J_3)}U(\hat{S})=\nonumber \\ H(-J_{1a},-J_{1b},J_{2a},J_{2b},J_3),\\ U(\hat{R})^\dag{H(J_{1a},J_{1b},J_{2a},J_{2b},J_3)}U(\hat{R})=\nonumber \\H(J_{1b},J_{1a},J_{2a},J_{2b},J_3),\eea
where $\hat{S}$ is the sublattice symmetry operation defined by $\hat{S}:\;\vec{S}(i,j)\rightarrow(-)^{i+j}\vec{S}(i,j)$ and $\hat{R}$ is a lattice rotation along the $a$-axis of angle $\pi$, that is, $\hat{R}:\;\vec{S}(i,j)\rightarrow\vec{S}(j,i)$. This fact tells us we only need to focus in the parameter region where $J_{1a}>|J_{1b}|>0$, and the phases in other regions can be obtained by applying the appropriate symmetry operations on the ground state within this parameter region. For example, if one has $J_{1b}<J_{1a}<0$, first we obtain the ground state for the Hamiltonian $H(-J_{1b},-J_{1a},...)$ and then transform the state under symmetry operation $\hat{S}\times\hat{R}$.

Second, as we have thoroughly discussed the phase diagram with $J_{1a,1b}>0$ in the previous paper \cite{Fang2009}, we only need to discuss the case where $J_{1a}>0$ while $J_{1b}<0$ in this note. This discussion is very simple because in this parameter region only one parameter, $J_{2b}$, is frustrated in the AFM3 phase, therefore the only possible instability is the one toward the ICB phase, the transition line of which is given by \bea J_{1a}-J_{1b}=4J_{2b}-8J_3.\eea The full phase diagram is given in Fig. S6.

We note that the effective exchange couplings are very close to the phase boundary between the $(0.5,0)$ AF ordered state and two different incommensurate magnetic states with wave vectors $(0.5-\delta,0)$ and $(0.5-\delta,0.5+\delta)$.  This is also consistent with the fact that the transition from the $(0.5,0)$ AF state to the incommensurate $(0.5-\delta,0)$ magnetic state is observed in Fe$_{1+y}$Te by increasing $y$ (refs. \onlinecite{Bao2009,Li2009}). There is also evidence that FeTe$_{1-x}$Se$_x$ can have $(0.5-\delta,0.5+\delta)$ incommensurate spin excitations \cite{Lumsden2010,Argyriou2010,Lee2010}.

\end{document}